\documentclass{./emulateapj}
\usepackage{color,epsfig,graphics,graphicx,./pdfpages,./psfig} 
\usepackage{natbib}
\definecolor{brown}{rgb}{0.6,0.4,0.2} 
\definecolor{purple}{rgb}{0.5,0,0.5} 
\shorttitle{Near-IR Spectroscopy of SN 2017eaw} 
\shortauthors{Rho et al.}

\def\msun{M$_{\odot}$}
\def\one{{\,\sc i}}
\def\two{{\,\sc ii}}
\newcommand{\iso}[2]{\ensuremath{^{#1}\rm{#2}}}
\def\nifs{\iso{56}Ni}
\def\cofs{\iso{56}Co}

\newcommand{\kms}{km~s$^{-1}$} 
\newcommand{\etal}{{et al.\/~}} 
 
\newcommand{\spitzer}{\textit{Spitzer}}

\shorttitle{Type II-P SN 2017eaw}

\begin{document}

\title{Near-Infrared Spectroscopy of SN 2017eaw in 2017:\\ Carbon Monoxide and Dust Formation in a Type II-P Supernova}

\author{ 
J. Rho\altaffilmark{1,2},  
T. R. Geballe\altaffilmark{3},
D. P. K. Banerjee\altaffilmark{4},
L. Dessart\altaffilmark{5},
A. Evans\altaffilmark{6}, and
V. Joshi\altaffilmark{4}
} 
\altaffiltext{1}{SETI Institute, 189 N. Bernardo Ave., Mountain View, CA 94043; jrho@seti.org}
\altaffiltext{2}{Stratospheric Observatory for Infrared Astronomy,
NASA Ames Research Center, MS 211-3,
Moffett Field, CA 94035}
\altaffiltext{3}{Gemini Observatory, 670 N. Aohoku Place, Hilo, HI, 96720, USA}
\altaffiltext{4}{Physical Research Laboratory, Navrangpura, Ahmedabad, Gujarat 380009, India}
\altaffiltext{5}{Unidad Mixta Internacional Franco-Chilena de Astronomia (CNRS UMI 3386), 
Departamento de Astronomia, Universidad de Chile, 
Camino El Observatorio 1515, Las Condes, Santiago, Chile}
\altaffiltext{6}{Astrophysics Group, Keele University, Keele, Staffordshire, ST5 5BG, UK}

{Accepted in ApJL}

\begin{abstract} 

The origin of dust in the early Universe has been the subject of considerable debate. Core-collapse supernovae (ccSNe), which occur several million years after their massive progenitors form, could be a major source of that dust, as in the local universe several ccSNe have been observed to be copious dust producers. Here we report nine near-infrared (0.8 $-$ 2.5~$\mu$m) spectra of the Type II-P SN 2017eaw in NGC 6946, spanning the time interval 22 $-$ 205 days after discovery. The specta show the onset of CO formation and continuum emission at wavelengths greater than 2.1~$\mu$m from newly-formed hot dust, in addition to numerous lines of hydrogen and metals, which reveal the change in ionization as the density of much of the ejecta decreases. The observed CO masses estimated from an LTE model are typically $10^{-4}$~M$_\odot$ during days 124 $-$ 205, but could be an order of magnitude larger if non-LTE conditions are present in the emitting region. The timing of the appearance of CO is  remarkably consistent with chemically controlled dust models of Sarangi \& Cherchneff.  
\end{abstract} 

\section{Introduction} 

Whether core-collapse supernovae (ccSNe) are a significant source of dust in the Universe is a long-standing question. The large quantities of dust observed in high-redshift galaxies \citep[e.g.][]{isaak02, watson15} raise a fundamental question as to the cause of its early presence, because stars, which are thought to produce most of the interstellar dust when on the AGB in the
modern Milky Way and nearby galaxies, could not have evolved to the dust-producing stage in very high-redshift galaxies. In contrast,
SNe from massive stars occur just millions of years after they form. Michalowski (2015) concluded that the early
appearance of dust could be explained if ccSNe do not subsequently destroy the dust they create. However, dust forms and grows in the interstellar medium \citep{spitzer78, zhukovska18}, and the dust in very high-{\textit{z}} systems might be produced primarily there, seeded by heavy elements and/or surviving dust from SNe \citep{draine09}. Depending on the progenitor mass, dust masses of 0.1--1.0 M$_{\odot}$ per ccSN \citep{todini01, nozawa03} can account for the amount of dust observed in very high-redshift galaxies.  Evidence that supernova-produced dust is the primary source of dust in the early Universe is still scarce \citep[see discussion in][and references therein]{gall11, rho18g54} and requires further study.

\spitzer\ mapping of the young supernova remnant (YSNR) Cas A revealed a remarkable similarity between the dust and CO distributions, confirming that dust forms and is preserved in at least some ccSNe ejecta \citep{rho08, rho12}. Four YSNRs including SN 1987A \citep{matsuura11,matsuura15}, the Crab Nebula \citep{gomez12, temim12, temim13}, Cas A \citep{rho08, deLooze17} and G54.1+0.3 \citep{temim17, rho18g54} have current dust masses of 0.1-0.9 M$_\odot$, consistent with theoretical models \citep{nozawa03, todini01}. The results suggest that some ccSNe are dust factories, and thus that they could be very important at high $z$, when other significant dust production processes may not exist. However, the fraction of ccSNe that produce this much dust is not known. Observing dust in additional local ccSNe and following its production and evolution in individual SNe over an extended period of time will help provide the answer. 

Current understanding of dust chemistry and composition in ccSNe is limited. Some models \citep[e.g.][]{nozawa03} suggest that dust forms 350 $-$ 900 days after a ccSN explosion, with carbon dust being one of the first condensates. Those models assume that little carbon is locked up in carbon monoxide (CO) and almost all of it is available for dust formation. CO is one of the most powerful coolants in the ejecta of type II SNe and, once it forms at several thousand K in some portions of the ejecta, is believed to be responsible in large part for cooling those ejecta to temperatures $\lesssim$~1700~K at which dust can form. Dust evolution models indicate that dust production depends significantly on the C/O ratio in the gas (Morgan \& Edmunds 2003; Dwek et al. 1998), as well as on the rate of destruction of CO, by impact both with the energetic electrons produced by the radioactive decay of $^{56}$Co and with He$^{+}$(Cherchneff \& Dwek 2010). Measurements of newly formed CO can shed light on the dust production rate.

\begin{table*}
\caption[]{Summary of Observations  and Results of CO Modeling}\label{Tobs}
\begin{center}
\begin{tabular}{cccccccc}
\hline \hline
& Day & exposure & CO mass & $T${\footnote{$\sigma$ $\sim$ 200 K;$^b$$\sigma$ $\sim$ 200 km s$^{-1}$; $^c$Photometry obtained by one of us (VJ) at 1.2-meter Mount Abu Infrared Telescope, India.; $^d$Spectra in Fig. 1 have been multiplied by these values.; $^e$The JHK imaging observations were taken on 2017 November 30.}}  & FWHM{$^b$} & $J$, $H$, $Ks${$^c$} & scale{$^d$}\\
      (yyyymmdd) &    & (sec) & (10$^{-4}$M$_{\odot}$)  & (K) & (km s$^{-1}$) & (mag) & factor\\
\hline
20170605 & 22 &  240   &  & ... & ... & ...  & 1.000\\  
20170616  & 33 & 360   & ... & ... & ... &11.61$\pm$0.04, 11.39$\pm$0.05, 11.16$\pm$0.04 &  0.505\\
20170626  & 43  & 360 & ... & ... & ... & ... &  0.218\\
20170706  & 53 & 360 & ... & ... & ... & ...  & 0.112\\
20170829  & 107 &  600 & $\leq$0.4 & ... & ... & ...  & 0.070\\
20170915  & 124 & 600 & 0.6--1.6  & 3000 & 2800 & ... & 0.051\\
20171001 & 140 & 1200 & 1.0--1.9 &  3300 & 2850 & ... & 0.012 \\
20171030  & 169  & 1200 & 1.6--2.2 & 3000 & 2850 &13.65$\pm$0.06, 13.45$\pm$0.08, 13.31$\pm$0.10& 0.005\\
20171205  & 205 &1080 & 1.9--2.2 & 2700 & 2750 &14.08$\pm$0.07, 14.02$\pm$0.10, 13.98$\pm$0.12$^e$ & 0.002 \\   
\hline \hline
\end{tabular}
\end{center}
\renewcommand{\baselinestretch}{0.8}
\end{table*}

\begin{figure*}
\begin{center}
\includegraphics[scale=0.6,angle=0,width=19.8truecm]{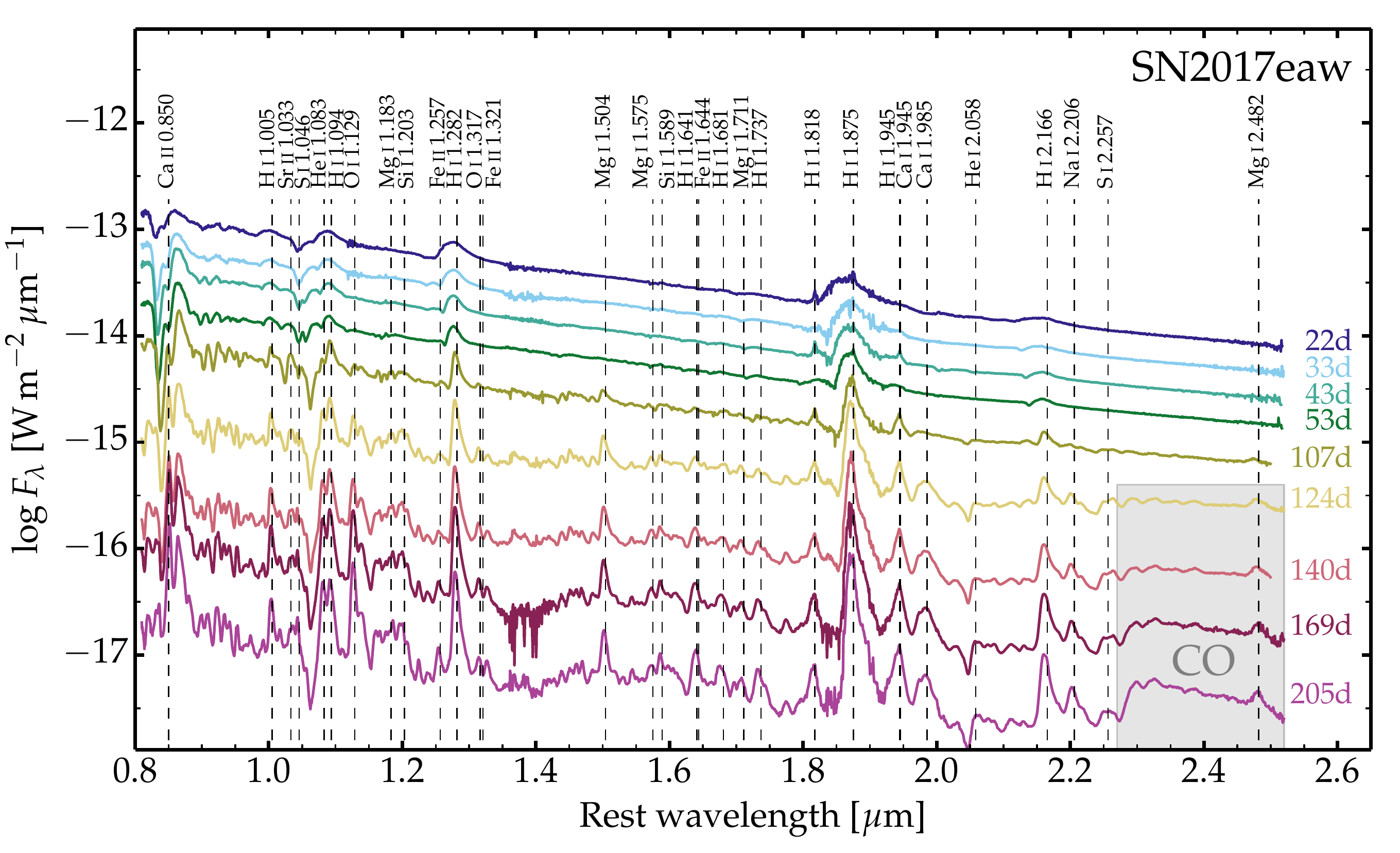}
\caption{
Gemini GNIRS 0.81 $-$ 2.52 $\mu$m spectra (dereddened) of SN 2017eaw obtained in 2017, in time order (top to bottom). Plotted spectra multiplied by values in Table~1. Prominent and important lines are labeled.  Wavelength interval with overtone CO emission is shaded. 
\label{fig1}}
\end{center}
\end{figure*}

Direct physical association of dust in high-{\textit{z}} galaxies with high-{\textit{z}} ccSNe is currently not possible, and thus to test the idea one must study dust production in local ccSNe. On 2017 May 14, a bright ccSN, 2017eaw, was discovered in the starburst galaxy NGC 6946. It has been classified as Type II-P \citep[ATel 10376;][]{xiang17}. The progenitor is known to be a red supergiant \citep[RSG; ATel 10378;][]{vanDyk17, kilpatrick18}. Because of its proximity and location in the sky, SN 2017eaw is ideal for studying the onset and evolution of both molecule and dust formation from a northern hemisphere telescope.  Here we present and describe nine Gemini near-infrared spectra obtained during 2017 June-December and discuss  the evolution of the spectrum, the detection of hot CO and an estimate of its mass, and emission from newly-formed dust.  ATel 10765 (Rho et al. 2017) is a brief report on the first six of these spectra. Estimates of the distance to NGC 6946 vary from 4 to 12.7 Mpc in the NASA/IPAC Extragalactic Database (NED). We adopt 7.72 $\pm$ 0.32 Mpc from \cite{anand18}.

 \begin{figure}
\includegraphics[scale=0.8,angle=0,width=8.7truecm]{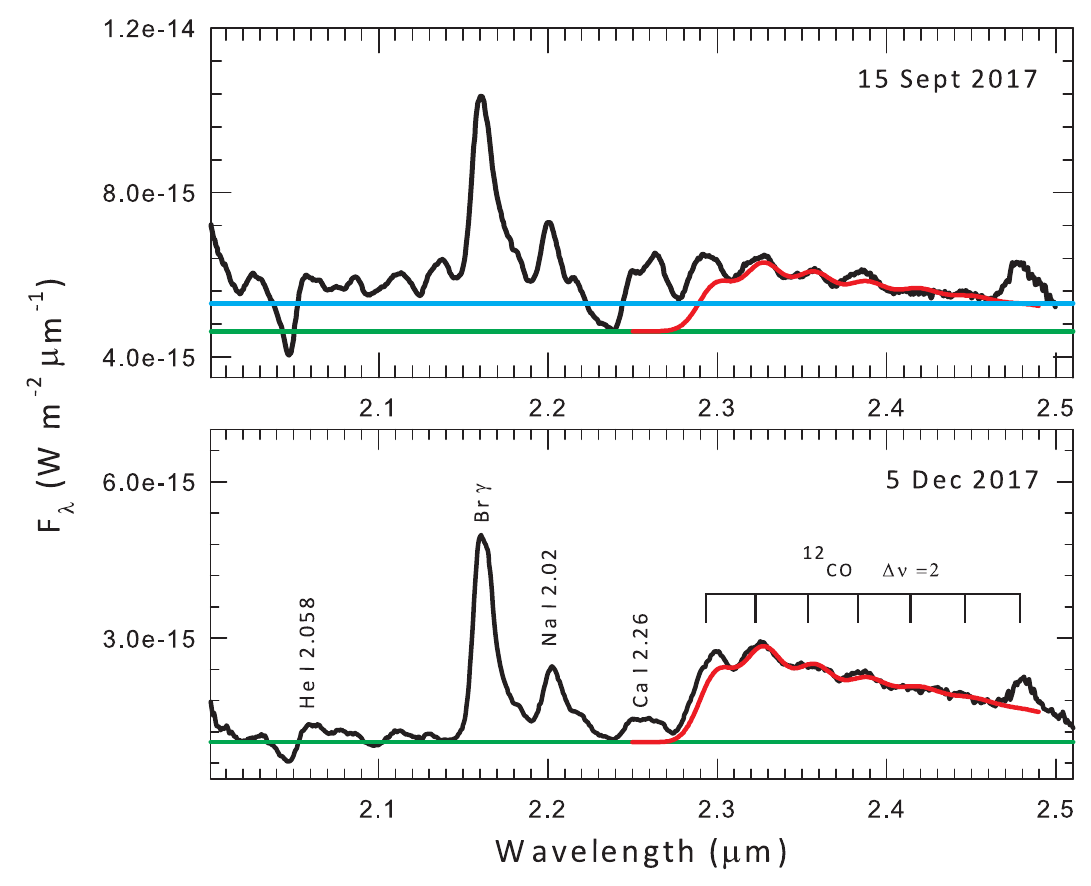}
\caption{
Zoomed in CO first overtone band emission superposed on LTE model fits (red lines) at 124~d and 205~d. Observed spectra in black. Green and blue lines in uppermost panel are possible continua. Parameters of fits are in Table 1.} 
\label{COmodel}
\end{figure}

\begin{figure}
\includegraphics[scale=0.6,angle=0,width=7.5truecm]{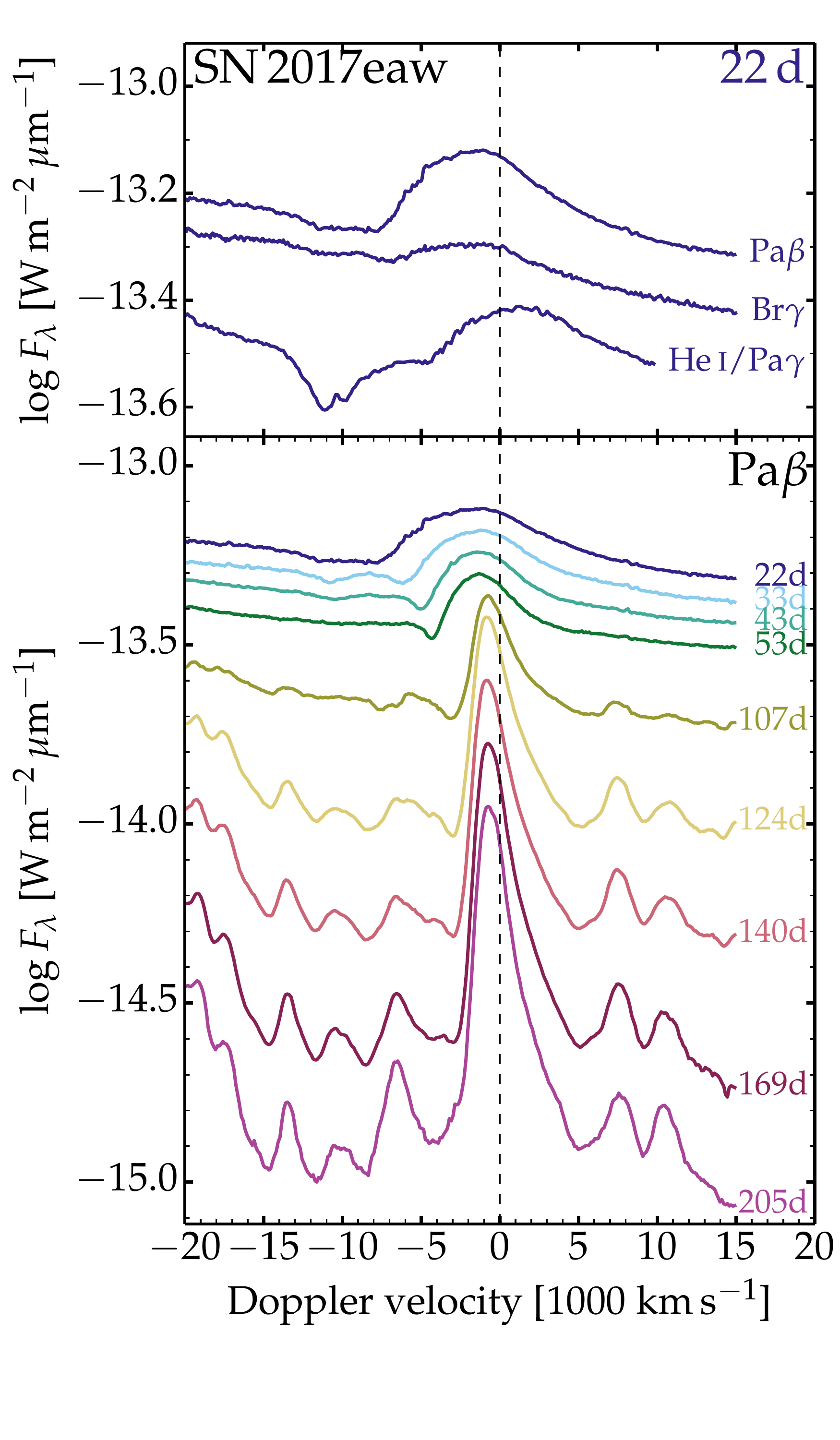}
\caption{Upper panel: velocity profiles of hydrgen Pa $\beta$ and Br $\gamma$ and blended H / He Pa~$\gamma$ / 1.083~$\mu$m on day 22. Doppler velocity relative to H\one\ lines. Lower panel: evolution of the Pa $\beta$ and Br $\gamma$ line profiles.}
\label{fig:profile}
\end{figure}

\section{Observations and Data Reduction} 

We obtained 0.81-2.52 $\mu$m spectra of SN 2017eaw using the Gemini Near-Infrared Spectrograph (GNIRS) on the 8.1-meter Frederick C. Gillett Gemini-North telescope, for programs GN-2017A-DD-8 and GN-2017B-DD-11. The observing dates and exposure times are listed in Table 1. GNIRS was configured in its cross-dispersed mode, using its 32 line/mm grating and a 0.45 arcsecond-wide slit to provide a resolving power of $\sim$1200 (250~km~s$^{-1}$). Observations were made in the standard stare / nod-along-slit mode with a nod angle of 3 arcseconds.  Nearby early A-type dwarfs, observed either just before or just after SN 2017eaw to minimize differences in airmass,  served as telluric and flux standards. Although the key questions we wished to address required only K-band spectra, and in particular the longer wavelength portion of that band, where hot newly formed dust and CO are both expected to emit, the full spectral coverage provides a wealth of information regarding the evolution of the kinematics, ionization state, and elemental abundances.

Data reduction utilized both the GNIRS cross-dispersed reduction pipeline and manual, order-by-order reduction for the shortest wavelength orders and for instances in other orders where the pipeline appeared to produce spurious results.  The manual reduction used standard IRAF{\footnote{IRAF is distributed by the National Optical Astronomy Observatories, which are operated by the Association of Universities for Research in Astronomy, Inc., under cooperative agreement with the National Science Foundation.}} and Figaro tools for flatfielding, spike removal, rectification of spectral images, extraction, wavelength calibration, and removal of hydrogen absorption lines in the spectra of the standard stars.  When order-by-order reduction was used, spectral segments covering different orders were stitched together after small scaling factors were applied, to produce final continuous spectra between 0.81-2.52 $\mu$m. The reliability of the spectra in the two intervals from 1.35 to 1.45 $\mu$m and from 1.80 to 1.95 $\mu$m is low because of low atmospheric transmission; one should view the spectra at these wavelengths with caution. In several of the spectra that were reduced manually the H\one\ lines in the telluric standard in the 1.80--1.95 $\mu$m interval were not removed prior to ratioing. At most other wavelengths, although the signal-to-noise ratio varies considerably, it is very high judging by the repetitiveness of the weaker features in spectra obtained on adjacent dates.  Because some narrow-slit  spectra were obtained in conditions of poor seeing and slightly non-photometric skies, flux densities coud be uncertain by $\pm$30\%.

\section{Overview of Spectra}

Figure 1 shows the nine 0.8--2.5~$\mu$m spectra of SN 2017eaw obtained during 2017. The earliest spectrum was obtained 22 days after discovery, the latest at 205 days. These data form the most extensive set of near-infrared spectra of a Type II-P supernova obtained to date. We are making them available to the astronomy community via WISEREP{\footnote{https://wiserep.weizmann.ac.il/}}.

The first four spectra, obtained during the photospheric phase, are dominated by hydrogen recombination line emission and the Ca\two\ triplet at 0.85~$\mu$m, all superimposed on a continuum that decreases monotonically with increasing wavelength.  The spectrum at day 107 marks a transition, as part of the ejecta begins to turn nebular in density and numerous neutral atomic lines begin to appear across the entire wavelength range. According to the optical photometry of \citet{tsvetkov18},  SN\,2017eaw turned nebular (i.e., the nebular brightness follows the 0.01\,mag\,d$^{-1}$ decay rate for \cofs) at an age of $\sim$125\,d.  At 124~d  the atomic lines in the near-IR spectra have become much more prominent, largely because the continuum has greatly decreased. 

Spectral signatures of several CO overtone bands at 2.3--2.5~$\mu$m are evident in emission in the 124 d spectrum and afterward (see Figures~\ref{fig1} and \ref{COmodel}), although the cut-on of the 2--0 CO emission near $\sim$2.29~$\mu$m is not yet obvious until 169 d. There is marginal evidence for weak CO emission earlier, at day 107. In addition, beginning on day 124 there is a flattening of the continuum at $\lambda$ $>$ 2.1~$\mu$m, not seen in the earlier spectra, which we interpret as emission from newly formed hot dust. The flattening of the continuum beyond 2.1~$\mu$m is also present in the final three spectra.

\section{Analysis and Discussion}
 
\subsection{Evolution of the Atomic Line Spectrum}

During the photospheric phase, in addition to the H\one\ and  Ca\two\ lines mentioned above, there are contributions from C\one, Mg\one, Mg\two, and later S\one\ (see Fig.~\ref{fig1}). Initially, the  H\one\ lines have P Cygni profiles and full extents of many thousands of km~s$^{-1}$ but the latter decrease monotonically with time as shown in Figure~\ref{fig:profile}. The extremely broad and shallow blue-shifted absorption troughs at the H\one\ Pa~$\gamma$,  Pa~$\beta$, and   Br~$\gamma$ ines in the early spectra may be the result of sustained absorption at high velocity by He\one, which results from an ionisation freeze-out of He in the outer ejecta \citep{D08_time}. Absorption by He\one\ lines at 1.083, 1.253, 1.278, 2.113, and 2.165~$\mu$m may contribute to these extended troughs.
 
In the observed nebular phase, the SN luminosity has dropped by two magnitudes since the photospheric phase, and a much higher fraction of the \cofs\ decay power emerges in lines. The nebular spectrum forms in the regions that absorb this decay power, located in what used to be the He core and the base of the H-rich envelope. While the  lines present in the photospheric phase (i.e. from H\one, He\one, Ca\two) persist, additional lines from intermediate mass elements (O\one, Na\one, Mg\one, Si\one, S\one, and Ca\one),  and iron-group elements (primarily Fe\one\ and Fe\two). 
Sr\two\,1.033\,$\mu$m are observed (see Fig.~\ref{fig1}), especially around the transition to the nebular phase. The line profiles in the nebular phase are narrower than earlier and the peaks in some of lines are blue-shifted by several hundreds of km~s$^{-1}$ (Fig.~\ref{fig:profile}). Early formation of dust, as in the case of SN2010ji which has similar velocity shifts \citep{gall14} could be the cause of this; two other possibile explanations are
continuum opacity of the ejected gas and asymmetric ejecta. 

We have generated a selection of simulations from \citet{dessart13} and \citet{dessart_fcl_18} for representative Type II SNe  from 15\,\msun\  progenitors, exploded to yield ejecta masses of 12.5 \msun\ with $1.2 \times 10^{51}$\,erg. For the photospheric phase the models are good matches to the observations. For example, the spectrum at 43\,d corresponds to a model of a red supergiant explosion \citep[model m15mlt3 from][]{dessart13} with a photospheric temperature of 5450\,K and a photospheric velocity of 4900 \kms. In the nebular phase, the overall agreement of models of blue supergiant explosion \citep[model Bsm from][]{dessart_fcl_18} endowed with 0.084\,\msun\ of \nifs\ and the spectra is satisfactory, although the models overestimate the strengths of Mg\one\ lines and underestimate the strengths of C\one\ lines around 1$\mu$m. The apparent shallow continuum $``$bump" observed at 1.40-1.75~$\mu$m (see Fig.~\ref{fig1}) is due at least in part to numerous overlapping lines, largely from Fe\one\ and Fe\two.

 \subsection{CO and Dust}
 
 \begin{figure}
\includegraphics[scale=0.8,angle=0,width=9.1truecm]{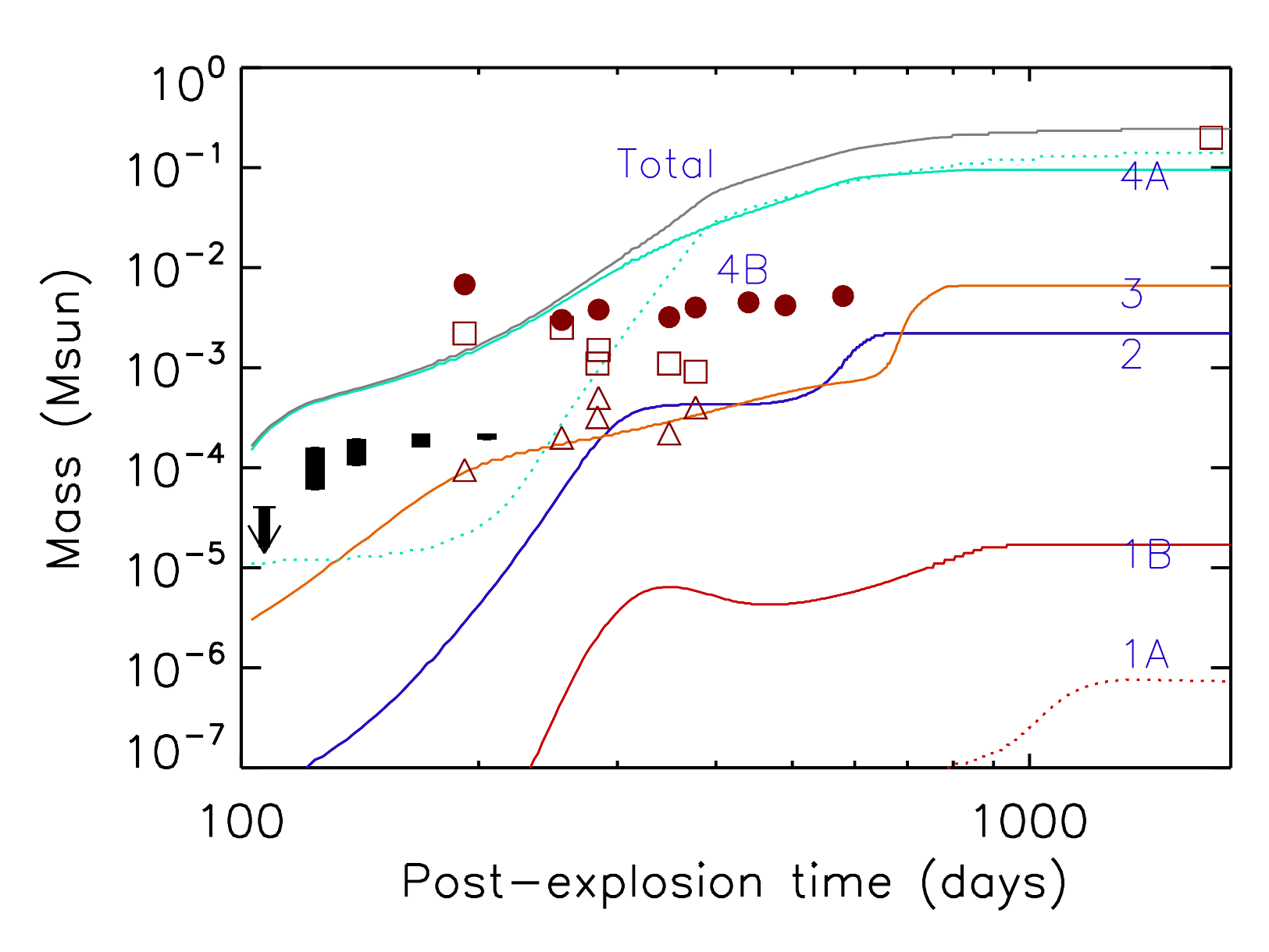}
\caption{CO masses of SN 2017eaw (black, filled rectangles) plotted on model of CO produced by a 15M$_\odot$ progenitor \citep{sarangi13}, for ejecta zones from innermost (1A) to outermost (4B). Vertical extents of black rectangles denote ranges of estimated CO masses, which depend on choice of 2.3--2.5-$\mu$m continuum. The zones include Si/S/Fe-rich (zone 1), O/Si/S-rich (zone 2), O/Mg/Si-rich (zone 3), O/C/Mg-rich (zone 4) \protect\citep{cherchneff09}.  Zones 1 and 4 are subdivided into A and B  subzones \citep[see Table 1 of][]{sarangi13}. The grey line represents the mass summed over all zones. The largest amounts of CO primarily form in zones 4A and 4B.  CO masses for SN 1987A derived for LTE  (small triangles) and non-LTE (small squares) from models of \citet{liu92} and small circles for thermal chemical model of \citet{liu95} are also plotted.}
\label{fig:CO}
\end{figure}

Carbon monoxide is first securely detected at 124 d. We have estimated the CO mass at days 124, 140, 169 and 205 using the LTE model developed by \cite{das09} and used by \cite{Banerjee16}, \cite{joshi17}, and others. Two examples are shown in Figure~\ref{COmodel}. The model assumes pure $^{12}$C$^{16}$O and optically thin, LTE emission. The presence of CO in each of the spectra is clear, based on the close match of peaks in the spectrum to the wavelengths of the 3--1, 4--2, and 5--3 band heads (2.323, 2.351, and 2.383 $\mu$m, respectively), the fair match to the 2--0 band head at 2.294~$\mu$m (which may be contaminated by an atomic line, probably C\one\ 2.291~$\mu$m) and, especially in the final two spectra, the large rise in flux density near the wavelength of the 2--0 band head. However, the strength of the CO emission and the CO mass derived from it are uncertain, due to the uncertainty  in the location of the K-band continuum, especially on days 124 and 140. For self-consistency, in each of the final four 2017 spectra we initially adopted a flat continuum (due to the photosphere and presumed dust emission) passing through the bottom of the trough at 2.23~$\mu$m. In doing so we essentially assumed that atomic line emission accounts for the elevated $``$pseudo-continuum" at 2.0--2.3~$\mu$m on those two dates, but does not contribute significantly at longer wavelengths where the CO emits. However, it is quite possible that these minima are P Cygni troughs.  If so, the above choice results in significant overestimates of the CO mass on days 124 and 140, but not on days 169 and 205, as on those dates the $``$pseudo-continuum" is much weaker than the CO emission. Thus, the upper limits to the derived CO masses listed in Table 1 for 124~d and 140~d should be regarded with caution. For example, if the adopted continuum in the top panel of Figure~\ref{COmodel} is the blue line, the CO emission strength on that date would be $\sim$3 times lower than that limit. The range of values of the CO mass in Table~1 reflect this uncertainty.

Best fit model spectra were determined by varying the CO mass, temperature, and velocity disperson with the goal of minimizing the reduced $\chi^2$ value over the region 2.30 to 2.46 $\mu$m (thus excluding the 2--0 band head and the strong [Mg~I] line at 2.48 $\mu$m). The fit results are summarized in columns 4-6 of Table 1. We obtain CO masses of $\sim$1$\times$10$^{-4}$ M$_{\odot}$ on all four dates from the LTE model and the assumed continua. The CO mass depends on the distance, $M_{7.72} (d/d_{7.72})^2$ where $M_{7.72}$ (in Table 1) and $d_{7.72}$ are the mass and distance assuming a distance of 7.72~Mpc, and $d$ is an actual distance to SN2017eaw. The masses on the first two dates may be considerably overestimated, as discussed above. On the other hand, if the molecular gas is not in LTE, the CO masses could be an order of magnitude larger, in analogy to the models of \citet{liu92} for SN 1987A. The best fit CO temperatures and velocity dispersions for all four spectra are typically 3000~K and 2800 km~s$^{-1}$, respectively. 

Observations of CO in SNe are fairly rare; detections from about a dozen SNe, including a Type Ic \citep{hunter09}, are known and  summarized  in \cite{sarangi18}. The evolution of the CO emission in ccSNe has only been studied over a lengthy time interval for SN 1987A (SNII-pec, a rare class but close to the Type II-P SN).  In Figure \ref{fig:CO} the estimated CO masses of SN 2017eaw are compared to those of SN 1987A and to the CO mass as a function of time predicted by \cite{sarangi13}. In both objects overtone CO emission is detected for the first time shortly after 100 days: 112 days for SN 1987A (Spyromilio et al. 1988) and at 124~d for SN 2017 eaw (implying significant CO production began between 107~d and 124~d). These dates are remarkably consistent with the chemically controlled dust models by Sarangi \& Cherchneff (2013, 2015), in which the first appearance of CO is predicted to be at $\sim$100 days. The  CO mass observed in SN 2017eaw ($\sim$10$^{-4}$M$_\odot$) is also roughly comparable to amounts in the models, in which most of the CO is produced in zones 4A and 4B \citep{sarangi13}. Individually each represents less than 0.1~\%\ of the available C and O masses (0.1 \msun\ and 1.0 \msun\, respectively, based on the models described above).  However, CO may continue to form during the first few or a few ten years after the explosion and thus may result in a vastly larger amount of CO than observed to be emitting in the first overtone band at any one time.

We have interpreted the flattening of the K-band continuum starting at 2.1~$\mu$m, in the four GNIRS spectra obtained from 124~d to 205~d, to be due to emission from hot newly condensed dust. When we fit the continua of the spectra above 2.1$\mu$m using carbon dust, dust temperatures of $\sim$1400 K and $\sim$1200 K are consistent with the spectra at 124~d and 205~d, respectively. We suggest that the carbon dust formed in SN2017eaw is graphitic carbon because it condenses at 1150 -- 1690 K, whereas amorphous carbon condenses at $\lesssim$1100 K \citep{fedkin10, ebel00}.  
Dust formation at 124 d is much earlier than predicted by some models  \citep{todini01, nozawa03}. However, the models of \cite{schneider04} and \cite{sarangi15}, and Lazzati and Heger (2015) suggest that dust formation begins at $\sim$150 d and continues until 2000 d. Recent observations show that dust formed at during 40--868 d in SN2010jl \citep{gall14} and continues to be present in SN1987A as late as 25 yr after outburst \citep{matsuura11, indebetouw14, zanardo14}.

The early sign of dust formation in SN2017eaw appears to be supported by an excess of 3.6-$\mu$m emission seen on day 193 in \spitzer\ IRAC observations (S. Tinyanont \etal\ 2018, private communication). This work serves as an early alert for further monitoring of the formation/evolution of the dust and CO in this object, particularly in the infrared. 

As of the date of the final spectrum (UT 2017 December 5), the rate of CO production remained strong. Because of this, we plan to obtain additional near-IR  measurements of SN 2017eaw during 2018. 

\acknowledgements
This paper is based on observations obtained at the Gemini Observatory, which is operated by the Association of Universities for Research in Astronomy, Inc., under a cooperative agreement with the NSF on behalf of the Gemini partnership: the National Science Foundation (United States), the National Research Council (Canada), CONICYT (Chile), Ministerio de Ciencia, Tecnolog'a e Innovaci—n Productiva (Argentina), and MinistŽrio da Cincia, Tecnologia e Inova‹o (Brazil). JR acknowledges support in part for this work from NASA ADAP grant (NNX12AG97G). Research at the Physical Research Laboratory is funded by the Department of Space, Government of India. We thank Dr. Arka Sarangi for providing his models, and the anonymous referee for insightful comments.

\bibliography{msrefs}

\clearpage

\end{document}